\documentclass[12pt]{iopart}
\usepackage[latin1]{inputenc}
\usepackage[T1]{fontenc}
\usepackage{graphicx}
\usepackage{amsfonts}
\usepackage{hyperref}
\usepackage{color}
\usepackage{soul}

\usepackage{cleveref}
\usepackage{etoolbox}
\crefname{equation}{}{}
\patchcmd{\numparts}{\addtocounter{equation}{1}}{\refstepcounter{equation}}{}{}

\begin{document}

\title{Equivalence between the Fitness-Complexity and the Sinkhorn-Knopp algorithms}

\author{D. Mazzilli$^1$, M. S. Mariani$^{2,3}$, F. Morone$^4$ and A. Patelli$^1$}
\address{$^1$ Enrico Fermi Research Center, via Panisperna 89a, 00184, Rome (Italy)\\}
\address{$^2$ Institute of Fundamental and Frontier Sciences,  University of Electronic Science and Technology of China, Chengdu 610054, P.R. (China)}
\address{$^3$ URPP Social Networks, University of Zurich, CH-8050 Zurich, (Switzerland)}
\address{$^4$ Department of Physics, New York University, New York, New York 10003, (USA)}

\ead{dario.mazzilli@cref.it, aurelio.patelli@cref.it}
\vspace{10pt}

\begin{abstract}
We uncover the connection between the Fitness-Complexity algorithm, developed in the economic complexity field, and the Sinkhorn-Knopp algorithm, widely used in diverse domains ranging from computer science and mathematics to economics. 
Despite minor formal differences between the two methods, both converge to the same fixed-point solution up to normalization.
The discovered connection allows us to derive a rigorous interpretation of the Fitness and the Complexity metrics as the potentials of a suitable energy function.
Under this interpretation, high-energy products are unfeasible for low-fitness countries, which explains why the algorithm is effective at displaying nested patterns in bipartite networks.
We also show that the proposed interpretation reveals the scale invariance of the Fitness-Complexity algorithm, which has practical implications for the algorithm's implementation in different datasets.
Further, analysis of empirical trade data under the new perspective reveals three categories of countries that might benefit from different development strategies.
\end{abstract}

\section{Introduction}
\label{sec:Introduction}
The Fitness and Complexity algorithm (FC) was first introduced in~\cite{Tacchella2012}, motivated by empirical regularities of international trades. 
Specifically, the most developed countries tend to produce and export most products, whereas developing nations export only a nested subset of such products~\cite{Tacchella2012}.
Therefore, the simple intuition behind the algorithm is that diversification is the signature of intangible capabilities necessary for the production of goods~\cite{Cristelli2013}.
Under this interpretation, each country cannot produce the commodities for which it does not own all the needed capabilities, and more complex products need higher levels of capabilities~\cite{Cristelli2013}.
The set of capabilities owned by countries could be measured, in principle, by collecting extensive data about their industrial systems and economies. Yet in practice, it is challenging to gather, harmonize and combine all the necessary information, and use it to operationalize diverse capabilities. 

Export data have allowed researchers to bypass this issue.
One can indeed infer the presence of the capabilities from a country's export basket, and capture the competitiveness of the country's productive system ("fitness") within a single variable~\cite{Cristelli2013}.
Indeed, under the assumption that countries produce and export most of the products they can within their capabilities, and that the products require different levels of capabilities, a natural hierarchy is detectable from the simple information of who produces and exports what.
Building on this idea, FC measures the Fitness of countries and the Complexity of products by only taking as input the binary country-product matrix $M_{cp}$, which accounts for the ability of country $c$ to export the product $p$.

Fitness and Complexity has been used to accurately predict the future development of countries~\cite{WorldBank2020,JRC2021}, and it plays a central role in the Economic Complexity framework to study and forecast macroeconomic systems~\cite{Tacchella2018}.
Compared to the Methods of Reflections~\cite{Hidalgo2009} and the equivalent Economic Complexity Index\footnote{\url{https://atlas.cid.harvard.edu/rankings}}, the fitness-complexity better quantifies countries' and products' structural importance in the country-product bipartite network~\cite{mariani2015measuring}; it substantially improves the GDP growth predictability~\cite{liao2017ranking}; it identifies correctly high-growth countries such as China and India, which are far from the top-ranked countries by ECI\footnote{For example, in 2000--2023, China has never entered in the top-15 countries according to the ECI [\url{https://atlas.cid.harvard.edu/rankings}, accessed 09.10.2023], which clashes with its unparalleled economic growth over the past 20 years. By contrast, China has been among the top countries by fitness-complexity, which indeed correctly predicts its growth in historical data~\cite{Tacchella2018}. In 2000--2023, India has never entered the top 40 by ECI [same link].}. Because of its effectiveness, the FC algorithm itself has seen modifications~\cite{mariani2015measuring,wu2016mathematics,Servedio2018} and applications in other domains~\cite{dominguez2015ranking,lin2018nestedness,Zaccaria2019,mariani2023ranking}. Compared to its variants such as the minimal extremal metric~\cite{wu2016mathematics} and the generalized fitness-complexity algorithm~\cite{mariani2015measuring}, the original fitness-complexity algorithm is substantially more robust with respect to noisy input data, which makes it better suited for the analysis of world trade data~\cite{mariani2015measuring,wu2016mathematics}.

Despite the central role of the fitness-complexity algorithm in the Economic Complexity field, several properties of the FC algorithm and its outputs are still unexplained. 
In particular, it remains unclear 
(1) why the FC is highly effective at rearranging $M_{cp}$ in such a way that a clear frontier separates a populated and an empty region of the matrix, see Results; 
(2) whether it is possible to derive the algorithm from optimization arguments;
(3) why for forecasting applications, it is more suitable to logarithmically transform the Fitness scores~\cite{Cristelli2013,Tacchella2018}; 
(4) how to rescale the algorithm to compare Fitness and Complexity scores across different datasets, e.g., international trade data across multiple years. 

Here we deepen our understanding of the Fitness and Complexity algorithm by connecting it with the long-known Sinkhorn-Knopp (SK) algorithm~\cite{Sinkhorn1967}. 
Originally introduced in the context of doubly stochastic matrices and matrix scaling~\cite{Sinkhorn1967}, the SK algorithm has been applied to pattern matching in computer vision~\cite{Peyre2019} and combinatorial matrix analysis~\cite{Brualdi2006}. 
Equivalent algorithms and variants have been applied in many fields under different names, such as Iterative Fitting Procedure and RAS method among others -- see~\cite{Idel2016} for an extensive review. 

We show that while FC and SK algorithms were introduced to solve different problems, their mathematical structure is the same. 
The main objective of this paper is to show the equivalence between the two algorithms, already hinted at by \cite{morone2022clustering}, discuss how it sheds light on the above-mentioned unexplained properties of the FC algorithm, and discuss additional implications for the Economic Complexity field.

\section{Methods}
\label{sec:Methods}
\subsection{Fitness and Complexity algorithm}
FC algorithm was developed by~\cite{Tacchella2012} and it is based on the definition of the Fitness of countries ( $F_c$ ) and of the Complexity of product ( $Q_p$ ) through the formulas:
\begin{numparts}	\label{eq:fit-com}
    \begin{equation}
		\tilde{F}_c^{(n)}=\sum_p M_{cp} Q_{p}^{(n-1)}
  \label{eq:fitness}
	\end{equation}
	\begin{equation}
		\tilde{Q}_{p}^{(n)}=\frac{1}{\sum_c M_{cp}\frac{1}{F_c^{(n-1)}}}
  \label{eq:complexity}
	\end{equation}
\end{numparts}
with a normalization step at the end of each iteration:
\begin{numparts}    \label{eq:fit-com-normalization}
    \begin{equation}
        F_c^{(n)}=\frac{\tilde{F}_c^{(n)}}{\left\langle\tilde{F}^{(n)}\right\rangle_c}
    \end{equation}
    \begin{equation}
        Q_p^{(n)}=\frac{\tilde{Q}_p^{(n)}}{\left\langle\tilde{Q}^{(n)}\right\rangle_p}
        \label{eq:renorm}
    \end{equation}
\end{numparts}
The matrix $M_{cp}$ is the binary representation of the bipartite network linking exporter country $c$ to exported product $p$.
This algorithm is iterative and converges to a fixed point of the coupled equations \ref{eq:fit-com} under some rather general conditions~\cite{Pugliese2016,wu2016mathematics}; researchers have introduced robust criteria to establish the convergence of the iterative equations~\cite{Pugliese2016,Servedio2018}.

The standard interpretation of Eq-\ref{eq:fit-com} is straightforward.
The Fitness of a country is defined as its diversification ($d_c=\sum_p M_{cp}$) weighted by the average Complexity of its exported products, such that a country exporting highly complex products obtains a large Fitness. 
In parallel, the Complexity of each product gets larger contributions from the countries with lower Fitness values thanks to the harmonic sum structure of eq.~\ref{eq:complexity}.

\subsection{Sinkhorn-Knopp algorithm}
The Sinkhorn-Knopp (SK) algorithm~\cite{Sinkhorn1967} (also called Bregman projection~\cite{Thibault2021}) is used to find the solution of the following problem: given a matrix $A \in \mathbb{R}_{+}^{n\times m}$ with non-negative entries, and two vectors $r \in \mathbb{R}_{+}^{n}$ and $c \in \mathbb{R}_{+}^{m}$ with non-negative numbers, find the two vectors $u$ and $v$ such that 
\begin{equation}
    \sum_j B_{ij} = r_i,\quad \sum_i B_{ij} = c_j,\quad B_{ij}\dot{=} u_iA_{ij}v_j.
\end{equation}
In other words, we want to transform the matrix $A$ such that its sums over rows and columns correspond to the given vectors $(r,c)$. 
In computer science, this is referred to as the matrix scaling problem~\cite{Idel2016}.
Of course, the constraint vectors must be balanced and sum to the same value, otherwise, a solution cannot be found.
The SK algorithm identifies $u$ and $v$ iteratively by computing
\begin{numparts}
    \label{eq:sink}
    \begin{equation}
        u_i^{(n)}=\frac{r_i}{\sum_j A_{ij}v_j^{(n-1)}}
    \end{equation}
    \begin{equation}
        v_j^{(n)}=\frac{c_j}{\sum_i A_{ij}u_i^{(n)}}
    \end{equation}
\end{numparts}  
The convergence of the algorithm is extensively discussed in the literature, see for example~\cite{Knight2008,Peyre2019} and references therein.
In general, matrix $A$ has to be non-negative with a total 
support~\footnote{For a square matrix to have total support 
means that for every positive element $A_{ij}>0$ there exists a 
permutation $\pi$ such that $A_{i\pi(j)}>0$ for all $j=1,...,N$.}~\cite{Brualdi1980,Berman2020}.
Notably, in the original work~\cite{Sinkhorn1967}, the second equation in~\ref{eq:sink} was written considering $u^{(n-1)}$ instead of its n-th iterate, but a later work~\cite{Anderson2010} found that the present formulation converges faster.

\subsection{Logarithmic barrier function}
The matrix scaling problems can be tackled by means of the logarithmic barrier function~\cite{marshall1968scaling}
\begin{equation}
    g(x,y) = \sum_{ij}x_iA_{ij}y_j - \sum_{i=1}^n r_i\ln x_i - \sum_{j=1}^m c_j \ln y_j
    \label{eq:potential}
\end{equation}
where $x$ and $y$ must be positive element-wise vectors and the logarithm assures the feasibility of the domain.
Any stationary point of $g$ solves $\sum_j A_{ij}x_iy_j = c_i$ and $\sum_i A_{ij}x_iy_j = r_j$ and solves the matrix scaling problem setting $u = x$ and $v = y$, and is locally marginally stable, as shown in~\ref{app:stability}.
Conversely, any matrix scaling solution gives a stationary point of $g$. 
This description highlights the nature of the logarithmic potential function of the vectors $u,v$. By comparing SK and FC we can say that $F$ and $Q$ are equivalent to $u,v$ of some logarithmic barrier function. This will be crucial in our new interpretation of the Fitness and Complexity indexes. It explains why F and Q are exponentially distributed and it will help us to interpret them as 'potential' of some form of energy, see Results and discussion.

\section{Results and Discussion}
\label{sec:Discussion}
%
\subsection{Correlation between algorithms}

The similarity between eq.~\ref{eq:fit-com} and eq.~\ref{eq:sink} becomes evident once one writes a symmetric version of Fitness and Complexity by transforming the Fitness via $X_c = 1/F_c$.
In fact, using $(X,Q)$ the two equations have exactly the same structure, although a few differences persist that could potentially lead to discrepancies.

First, $F$ and $Q$ in eq.~\ref{eq:fit-com} at stage $n$ are both updated using the values at step $n-1$. 
Indeed, with this prescription, FC generates two different series of values, one for even and one for odd $n$, both converging to the same fixed point.
On the other hand, SK updates the two values sequentially, leading to a single chain of convergent values.
Second, FC does not consider the vectors $(r,c)$, implicitly setting their entries to unity.
However, $r_i=1$ and $c_j=1$ are not feasible in SK because their sums must be equal in order to balance the constraints and this is not possible unless the matrix $A$ is square.
Third, the normalization of eq.~\ref{eq:fit-com-normalization} is considered by FC in order to guarantee finite values and more importantly to re-balance F and Q, while SK does not need it.

Given these differences, one may wonder if the two algorithms lead to the same result, upon setting $r$ and $c$ to a constant value (such that their sum are equal). 
We test this hypothesis by applying SK and FC to the export data, a set of data widely considered in economic complexity~\cite{Hidalgo2009,Tacchella2012}.
Whilst not being numerically exactly the same, the ranking of $F$ and $u$ are perfectly matching and their Pearson correlation coefficient is basically 1 within the numerical precision of a float ($10^{-8}$).
Therefore, the two approaches give the same solutions, regardless of the actual algorithm implemented.
Balancing the equations on one side and implementing the normalization on the other might be both effective ways to find the fixed point.

\subsection{The potential interpretation of Fitness and Complexity}
The reinterpretation of the FC in terms of potentials of the logarithmic barrier function sheds light on some properties of $F$ and $Q$ that were previously unclear.

\paragraph{Fitness' logarithmic scale.}
Function $g$ highlights that the natural scale of the fitness score is the logarithmic scale, as often considered in many diagrams (for example, the Income-Fitness diagram in~\cite{Tacchella2012}), because the problem corresponds to a constrained optimization problem with logarithmic constraint barrier.
In fact, for relevant applications (see for example~\cite{WorldBank2020,JRC2021}), a logarithmic transformation is required in order to have well-separated values of Fitness and Complexity, as for example in the visualization of countries' trajectories, see below.

\begin{figure}[t!]
	\centering
	\includegraphics[width = \textwidth]{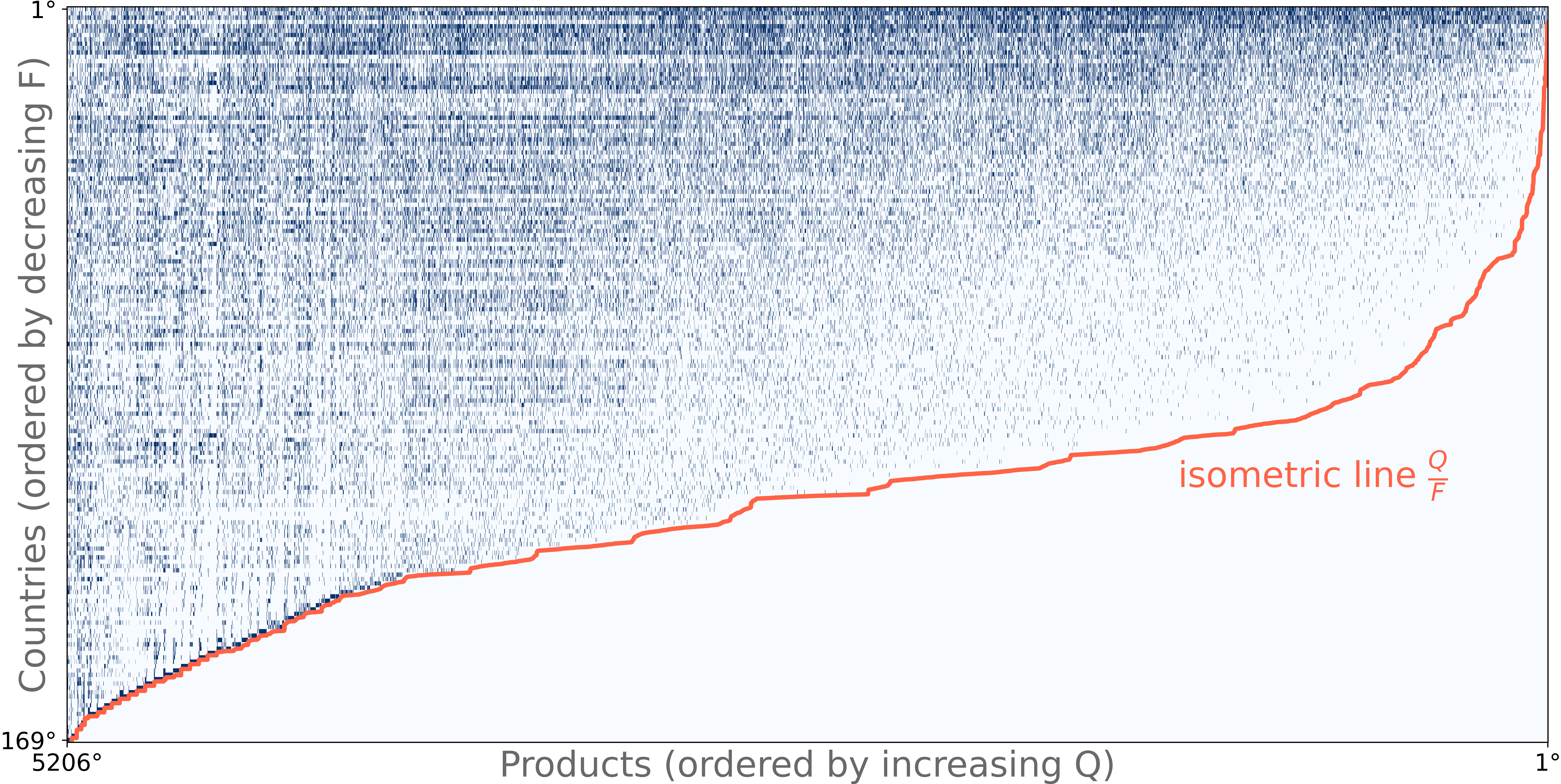}
	\caption{Matrix representing the bipartite network of the export flow in 2016, reordered using the ranks of Fitness and Complexity. The red line represent the constant-line of Complexity over Fitness at the border.
	The network is based on the International Trade data using the Harmonized System classification HS-2012, the golden standard for the year 2016.
	}
	\label{fig:matrix_export}
\end{figure}

\paragraph{Potentials.}
Further, a well-known property of FC is its ability to reorder rows and columns of $M_{cp}$ in order to obtain a upper-left \textit{triangular} matrix which highlights its nestedness~\cite{Mariani2019}, as shown in figure~\ref{fig:matrix_export}. 
This can be obtained by simply reordering rows and columns of the matrix by the ranking of $F$ and $Q$, and this feature can be explained in terms of potentials functions (also called Kantorovich potentials). 
The Fitness of a country measures its industrial system's capability while the Complexity is a measure of how hard is to produce each commodity. 
Thus, the region of the matrix $M_{cp}$ with high $Q$ and low $F$ should not be populated, by construction, because that would mean low Fitness countries producing highly complex products.
Indeed, the two logarithmic potentials in eq.~\ref{eq:potential} define a barrier that cannot be overcome: here is a value of $\ln x + \ln y$ whose constant-value line defines the separation between the feasible and forbidden regions of the matrix. 
Translated in terms of Fitness and Complexity, we can find a value of $Q/F$ that defines a barrier line in the reordered shapes that meets the same requirements with striking precision, see fig.~\ref{fig:matrix_export} where we plot the isometric line with the lowest value among all the lines that define a completely empty region under the curve. 
Operationally, the red line in the figure is found by trying different values of $Q/F$ and checking if any non-zero entries of the matrix fall below the curve.

This explains why in many real systems, by rearranging the matrix's rows and columns by FC, the parts of the rearranged matrix exhibit a continuous border line.

Taking further our interpretation, the sum $\sum_p M_{cp}Q_p/F_c$ corresponds to the constant vector $r_c = 1$, up to a scale (see below), as assumed in FC. Let us now interpret $Q_p/F_c$ as the amount of energy associated with the production of $p$ by country $c$.
Under this interpretation, constraint $r_c$ represents the \emph{total energy} possessed by country $c$, and FC assumes (albeit unintentionally) fair availability of energy resources among nations. 
The fairness assumption is enforced by the condition that $r_i$ is homogeneous across countries, which makes the FC algorithm independent of country-level properties other than their export basket composition.  
Every country allocates resources into its productive system by distributing them among the production of different products; this information is encoded in $M_{cp}$.  
On the other hand, the ability of different countries to produce a product is not fairly distributed: at different levels of complexity $Q$, the energy cost differs.
This means that each given resource unit can be exploited with different efficiency by countries with different Fitness $F$, and the energy cost changes accordingly to the ratio of the two metrics. 

It is valuable to define matrix element $E_{cp} = Log(Q_p/F_c)$, which caputres the "energy cost" of element $M_{cp}$. It should be clear to the reader that this is potential energy, hence it is relative to an offset. A complexity $Q=1$ does not mean that the corresponding product can be exported effortlessly. To give a numerical example, a product of complexity $Q = 1.4$ would cost $E = 2.64$ for a low Fitness country with $F=0.1$ and just $E=-1.46$ for a high fitness country with $F = 6$.
Therefore, the forbidden region of $M_{cp}$, displayed in fig.~\ref{fig:matrix_export}, is simply represented by too-expensive products in terms of resource allocation. 
The heterogeneity of resources owned by countries can be naturally introduced in this framework in further studies, see conclusions.

\paragraph{Scale invariance.}
Another feature that FC inherits from the logarithmic barrier function is that $g$ presents a symmetry and, as a consequence, the stationary solutions of equation~\ref{eq:potential} are not unique.
The transformation $(x,y)\rightarrow (x/\alpha,y\alpha)$ with $\alpha\in\mathbb{R}_+$ leaves invariant the value of $g$, which allows a \textit{rescaling} of the fixed points.
At the level of the numerical implementations of the algorithm, the rescaling helps to increase the numerical precision and consequently the stability of the codes~\cite{Thibault2021}.
Instead, on the conceptual ground, the rescaling symmetry states that the fixed point solutions are not uniquely defined, rather a scale must be chosen to break the symmetry.
The scale becomes particularly relevant when we are interested to compare the Fitness scores derived from different matrices, such as when comparing the Fitness scores of the same countries at two different years.

There are different ways to set a scale. For example, it is possible to select a single country or a product as a reference, as proposed by~\cite{Patelli2021}. 
Another possibility to define a scale, which we introduce here, is to add a dummy country able to produce all the products, thus adding a full row of ones to the $M_{cp}$ matrix.
The value at the fixed point of the dummy country's Fitness sets the scale by requiring that at any time its value is fixed to a constant (\emph{e.g.} 1). Setting the scale provides a formal method to compare different matrices, and it affects only the Fitness scores and trajectories and not the ranking.

\begin{figure}[t!]
	\centering
	\includegraphics[width =0.475\textwidth]{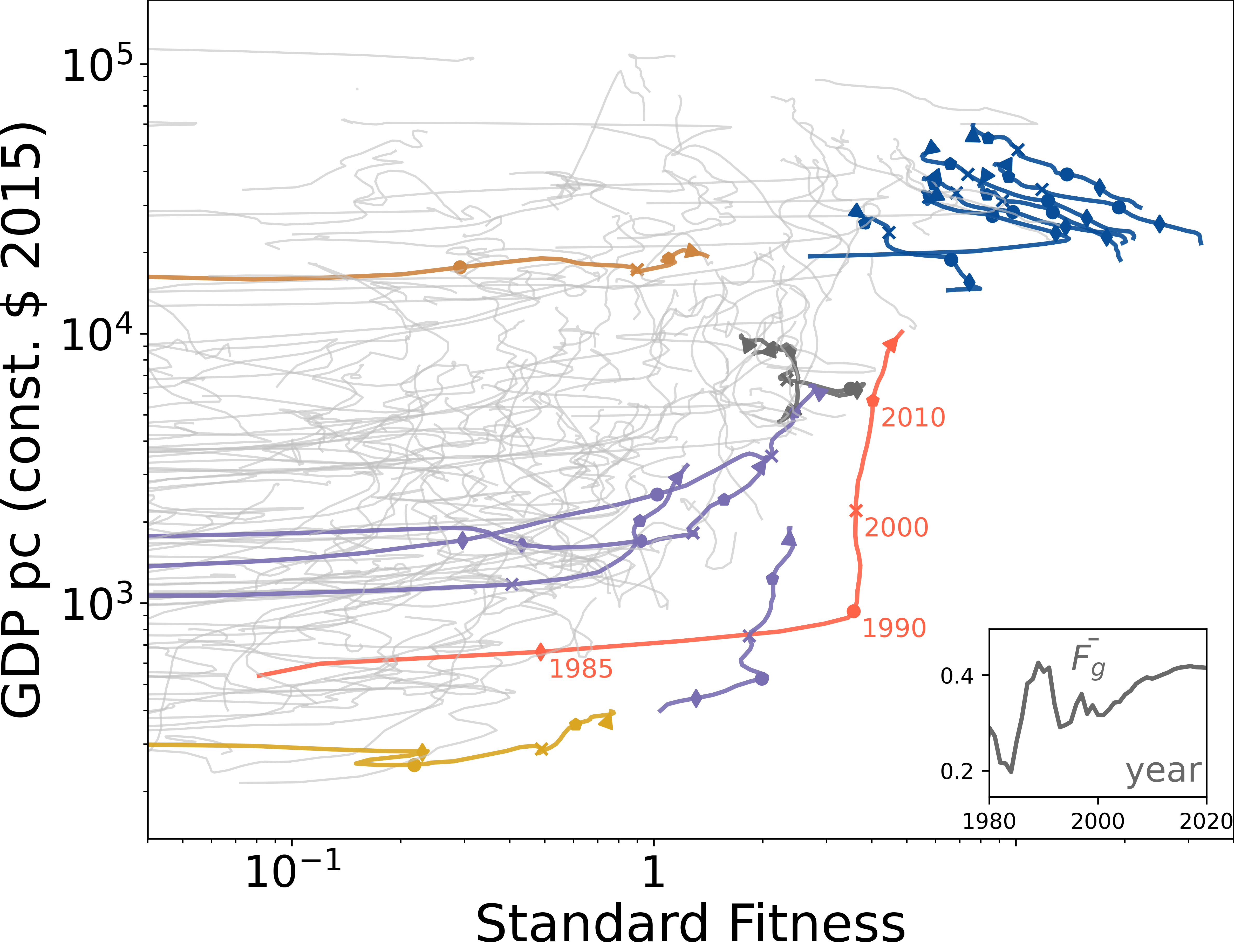}
	\includegraphics[width =0.475\textwidth]{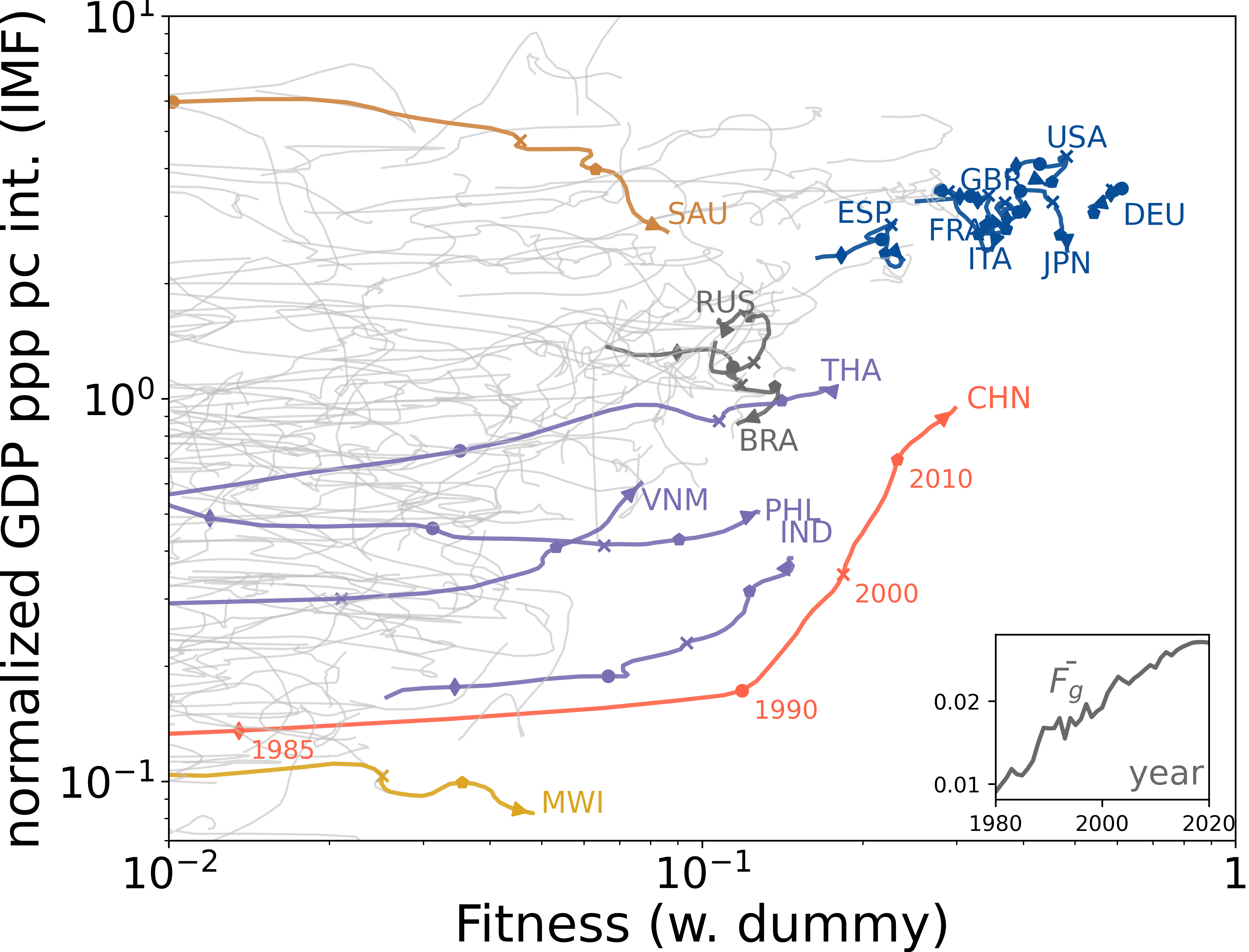}
\caption{Trajectories of different countries in the Income --- Fitness plane. We compare the standard Fitness (left panel) and the Fitness evaluated adding the dummy country (right panel). The insets show the average Fitness, indicating that there is a general trend. The Fitness are computed using the SITC v2 classification, which dates back to 1976.}
	\label{fig:trajectories}
\end{figure}

Using the International Trade export (see Data), we verify that the trajectories in the Income - Fitness diagram can be substantially affected by the choice of the Fitness' scale.
Figure~\ref{fig:trajectories} shows in the right panel that when we fix the Fitness score of the dummy country, the Fitness of Western countries (blue lines) does not evolve appreciably over more than 4 decades, whereas many Asian countries (purple lines), and China on top (red line), are catching up the wealth and developed nations.
Instead when the scale is fixed by the normalization condition as in the original paper~\cite{Tacchella2012}, we observe that the Fitness of wealthy countries decreases in time, as shown in the left panel.
The same effect can be observed in the Income variable, since removing the global income average, make stationary most of the wealthy countries.
The main takeaway from these results is that while ether scale-invariance of potential $g$ allows the researcher to set the scale, this choice should be made cautiously, as it affects the resulting fitness temporal trajectories and their interpretation.

\begin{figure}[t!]
    \centering
    \includegraphics[width = \textwidth]{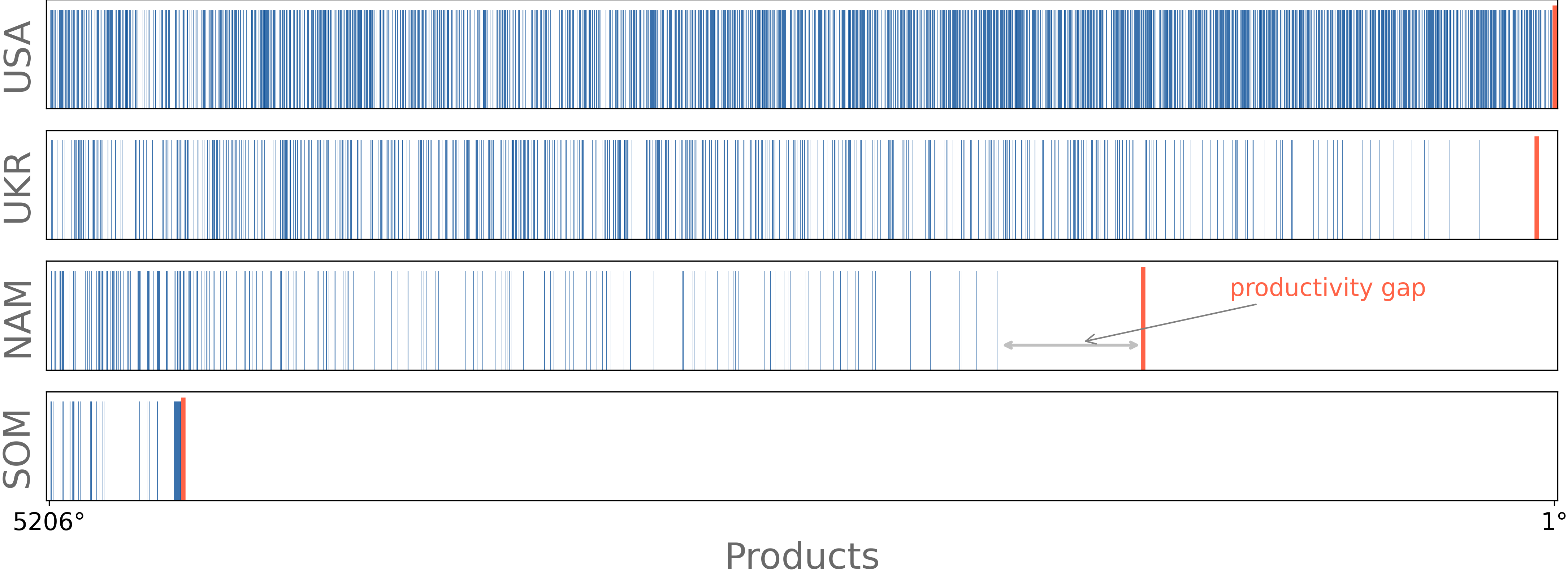}
    \caption{Spectra of exports in 2016 of some representative country: USA, Ukraine (UKR), Namibia (NAM) and Somalia (SOM). Each blue line indicates that the country is actively producing the associated product. The products are disposed on the horizontal axis are ordered following the rank of Complexity. The red line indicates the position of the border line of figure~\ref{fig:matrix_export}.}
    \label{fig:spectra}
\end{figure}

\paragraph{Possible development pathways.}
The existence of a potential barrier makes it also possible to identify Fitness as the maximum \textit{sophistication} achievable by products exported from each country, which is represented by the isometric line of the $M_{cp}$ matrix. However, most of the developed countries do not export products lying on the isometric line (see Fig.~\ref{fig:matrix_export}), and exhibit a lower density of exported products near the isometric line, which indicates a productivity gap with respect to the maximum product-Complexity level achievable from their Fitness.
This pattern suggests that different nations might benefit from pursuing different strategies to increase their capabilities and ultimately their wealth.
We conjecture below that different countries might benefit from three different classes of new product development strategies: 
\begin{itemize}
    \item \textbf{Learner pathway,} which might be suited for low-Fitness countries that are at their maximum level of capabilities and already reached the isometric line. A clear example is Somalia in Fig.~\ref{fig:spectra}. These countries might benefit from enhancing their industrial system by increasing the relevant capabilities before entering new markets. 
    
    \item \textbf{Exploiter pathway,} which might be suited for medium-Fitness countries, such as Namibia and Ukraine, whose spectra of exported products~\cite{operti2018dynamics} lie far from the isometric line or who exhibit low-density of exports near the line. These countries may have unexpressed capabilities, and could benefit from investing in the production of goods already allowed by their capabilities. 
    
    \item \textbf{Explorer pathway,} which might be suited for high-fitness countries that present a dense spectrum along the entire axis (see USA in the figure). These countries might be already efficiently exploiting their capabilities. Therefore, their best strategy might be to move from the exploitation of existing capability to the exploration of new, and thus not classified, activities and goods.
    
\end{itemize}

\begin{figure}[htbt]
	\centering
	\includegraphics[width =0.475\textwidth]{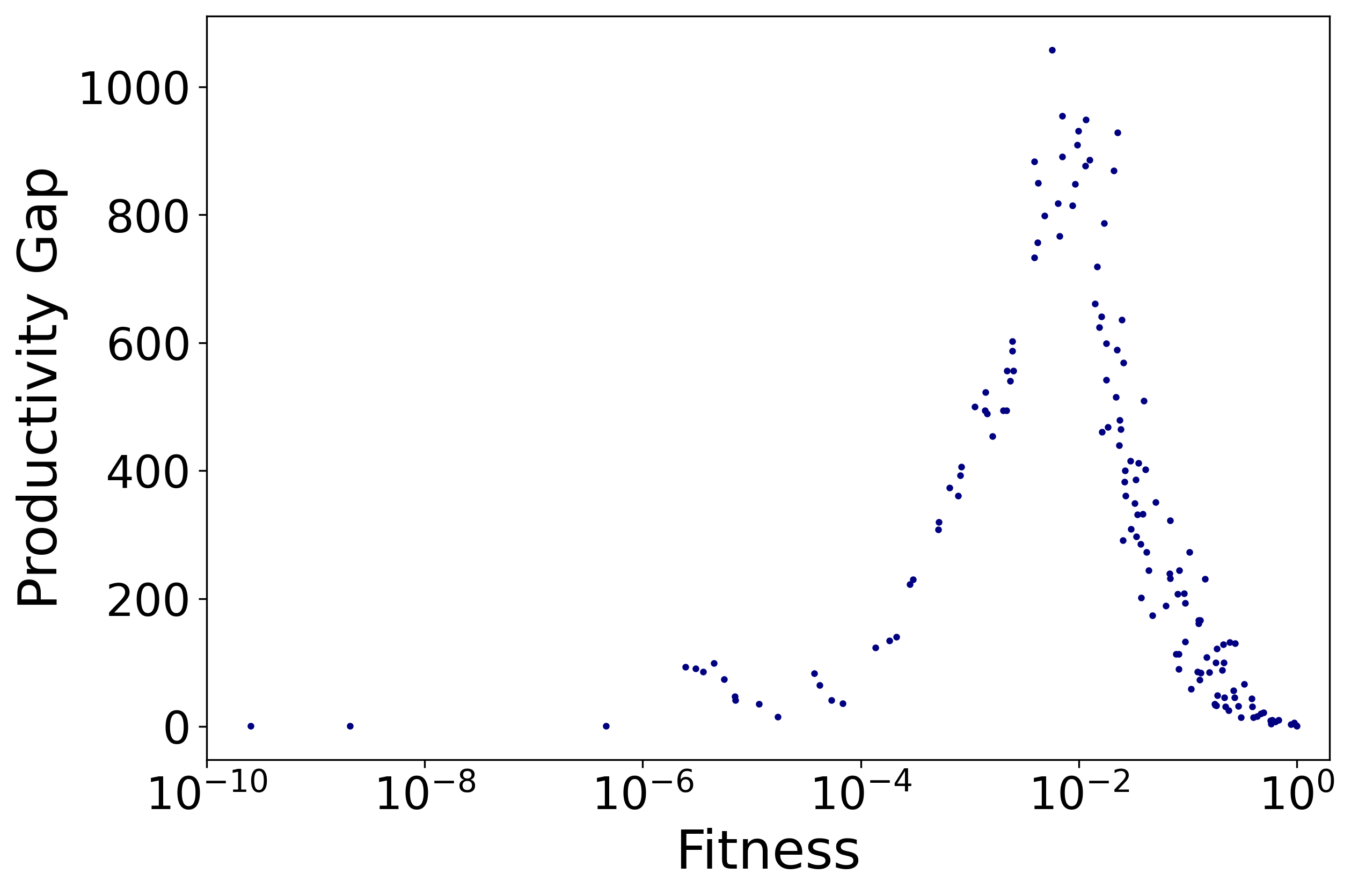}
	\includegraphics[width =0.475\textwidth]{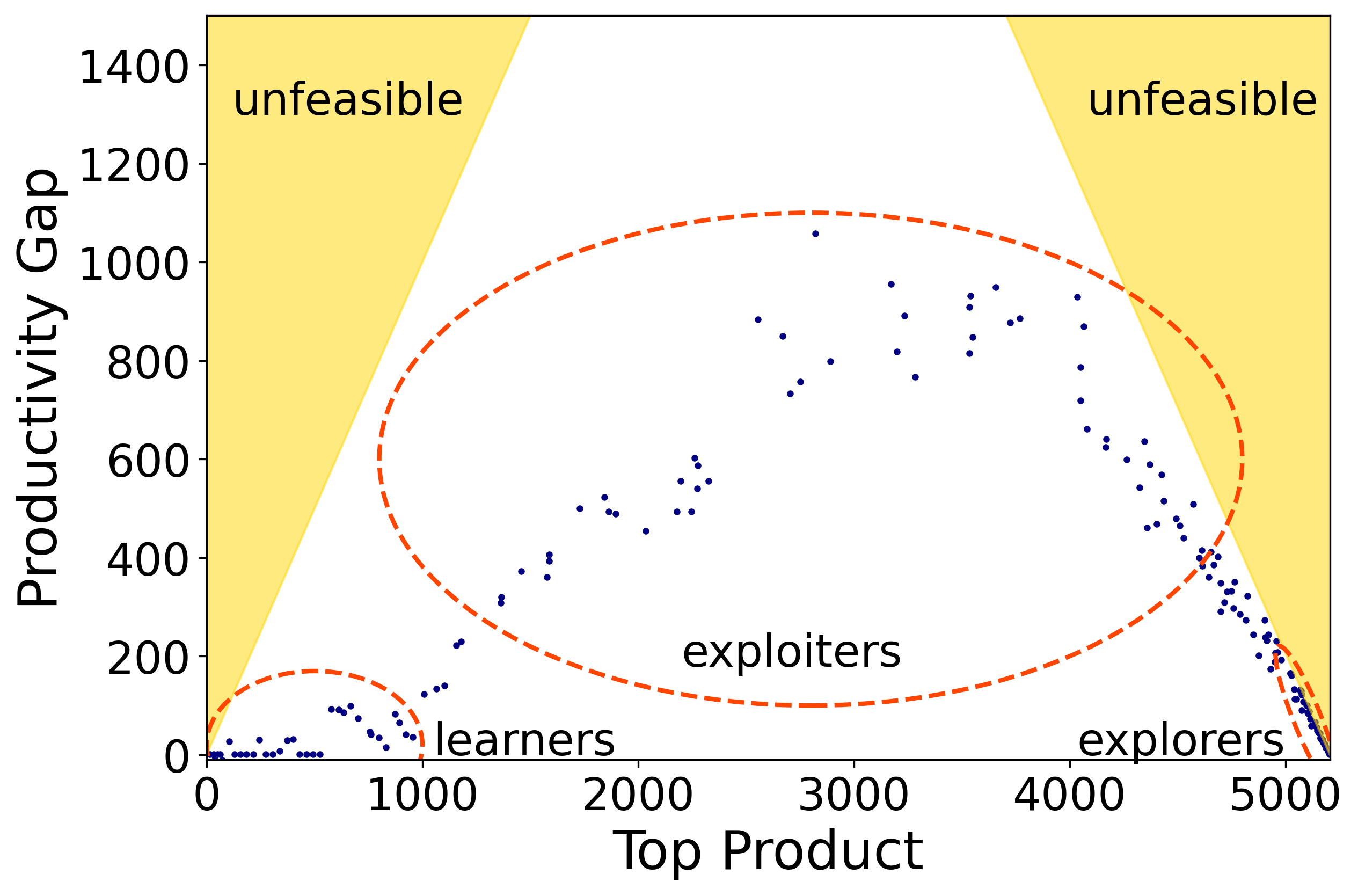}
\caption{Left panel: scatter plot Fitness Vs Productivity gap. Right panel: classification of development pathways in the plan Productivity gap - Top exported product.}
	\label{fig:prod_gap}
\end{figure}

The productivity gap is clearly related to the Fitness and the top exported product (by complexity), as shown in Fig.~\ref{fig:prod_gap}. 
The left panel shows an intriguing U-shaped relationship between countries' fitness and productivity gaps: It is not the highest-fitness or the lowest-fitness countries that exhibit the largest productivity gaps, but the countries with intermediate fitness values. Intermediate-fitness countries are likely those with unexpressed capabilities, that might benefit the most from the exploitation pathway described above.
The three categories of countries can be loosely identified in Fig.~\ref{fig:prod_gap}, right panel. The countries that might benefit the most from the learner and explorer pathways are close to the forbidden region of $M_{cp}$, whereas those that may benefit the most from the exploiter pathway exhibit middle top-product complexity, intermediate fitness values, and a very large productivity gap. In appendix~\ref{app:gap-table} we show the entire list of countries with their productivity gap and top exported product.

Clearly, Fitness is a single parameter and cannot fully explain the productivity gaps.
A finer description of export baskets is needed to identify opportunities for new products and optimal development strategies, which we leave for future studies.
For example, our simplified classification could be integrated with network-based approaches~\cite{Pugliese2017} and machine learning approaches~\cite{Tacchella2021} to better predict product development probabilities, which might suggest new future development pathways for countries.

\section{Conclusions}
\label{sec:Conclusion}
We showed the deep connection between the Fitness and Complexity algorithm, developed in Economic Complexity, and the Sinkhorn-Knopp algorithm devised in matrix scaling. 
Indeed, despite minor differences in the algorithms' structure, the fixed points of the two algorithms coincide. 
This connection allowed us to better understand some properties of FC and interpret Fitness and Complexity as potentials of a suitable energy function of mutualistic systems such as the international trade web.
This description indicates how countries deploy their units of resources, it allows the definition of a barrier that defines the inaccessible region of country-product matrix, and it helps us to classify countries by their exploited potential, which suggests different strategies of investment and development. 
This classification is to be intended as a broad description of the global competition in international trade. Further studies will test the significance of these categories by looking in detail at their productive systems and their evolution through time. 

The use of a dummy country to set a common scale across different $M_{cp}$ finds immediate application in all numerical employments of Fitness and Complexity, from GDP forecasting~\cite{Tacchella2018} to studies of relatedness of product~\cite{Tacchella2021}. In principle, all studies that analyse the dynamics of these indicators should adopt a way to fix the scale before comparing datasets of different years, sizes or sources.

Arguably, the most interesting follow-up of this work is the connection it represents with the Optimal Transport theory. SK algorithm is an efficient way to solve the Optimal Transport problem~\cite{Cuturi2013}. Thus, with this work, we opened a door to a vast and rich theory that can be used to extend, interpret and enhance the economic complexity framework. For example, differences in resources and market demands can be accounted for by framing a resource allocation problem equivalent to a classic transport optimization.

\section{Data}
Export data are from COMTRADE \footnote{https://comtradeplus.un.org/}. Only for figure 2 we use data based on the SITC v2 classification started in 1976 to have a long time series, needed to show the regularization effect on trajectories. For all other results and plots we use Harmonized System (HS-2012), reconciled by the procedure described in~\cite{Tacchella2018}, starting in 2012 and ending in 2021.
Despite the SITC classification has a longer scope, the quality of the results is less accurate for two main reasons.
On one hand, SITC v2 is based on a rigid classification defined in the 70s, thus missing many products that did not exist at the time.
On the other hand, SITC has a definition of roughly 200 codes of products, while HS has more than 5000 at the 6-digit depth considered in the work.

\section{Acknowledgement}
FM was partially supported by AFOSR (Grant No. FA9550-21-1-0236).
MSM was supported by the URPP Social Networks of the University of Zurich and the Swiss National Science Foundation (Grant No. 100013-207888).

\appendix
\section{Stability of the fixed point}
\label{app:stability}
The stability of the fixed point can be evaluated better by focusing on the potentials, a.k.a. the logarithms of $z_i=(x,y)_i=e^{\varphi_i}$.
The logarithmic barrier function reads
\begin{equation}
    g(\varphi) =\sum_{ij} e^{\varphi_i}e^{\varphi_j}M_{ij} - \sum_id_i\varphi_i
\end{equation}
where the matrix $M$ and the constraint $d$ depend on the previous terms through
\begin{equation}
    M_{ij} = \frac{1}{2}\left(\begin{array}{cc}
		0 & A \\
		A^T & 0
		\end{array}\right), \quad d=(r,c).
\end{equation}
The Hessian at the fixed point ($\varphi^*$ s.t. $\partial_{\varphi_i}g(\varphi^*)=0$) can be written as
\begin{equation}
    \mathcal{H}_{ij}= \partial^2_{\varphi_i\varphi_j}g(\varphi^*) = \delta_{ij} d_i + e^{\varphi^*_i} e^{\varphi^*_j}(M_{ij}+M_{ji})
\end{equation}
The Hessian matrix $\mathcal{H}$ is a symmetric diagonally dominant matrix with real non-negative diagonal entries, hence it is positive semidefinite~\cite{horn_johnson_2012}.
Therefore, the fixed-point solution is marginally stable, since we already know and discuss in the main text that there is at least one direction of constant energy related to the scale invariance symmetry.

\newpage
\section{Countries' productivity gap}
\label{app:prod-gap}
\begin{table}[h!]
\scriptsize
\begin{tabular}{|lrr|lrr|lrr|}
\hline
Country &  Top product &  Gap & Country &  Top product &  Gap & Country &  Top product &  Gap \\
\hline
 DEU &         5205 &            1 &  SRB &         4903 &          273 &  PYF &       2326 &        556 \\
 CHN &         5204 &            2 &  AUS &         4875 &          201 &  BLZ &       2277 &        587 \\
 JPN &         5203 &            3 &  SLV &         4852 &          244 &  SLE &       2275 &        540 \\
 GBR &         5202 &            4 &  LBN &         4823 &          322 &  BOL &       2262 &        602 \\
 USA &         5200 &            6 &  CHL &         4817 &          273 &  BRN &       2247 &        494 \\
 FRA &         5198 &            8 &  MAR &         4790 &          285 &  MNG &       2197 &        556 \\
 NLD &         5197 &            9 &  ARG &         4765 &          351 &  OMN &       2180 &        494 \\
 IND &         5196 &           10 &  MKD &         4759 &          297 &  ZMB &       2035 &        454 \\
 ITA &         5196 &           10 &  COL &         4749 &          332 &  RWA &       1895 &        489 \\
 CHE &         5189 &           14 &  JOR &         4730 &          331 &  HTI &       1863 &        494 \\
 POL &         5188 &           14 &  CRI &         4718 &          309 &  MWI &       1843 &        523 \\
 KOR &         5187 &           16 &  ARE &         4700 &          349 &  AFG &       1729 &        500 \\
 ESP &         5183 &           20 &  KEN &         4699 &          291 &  TJK &       1588 &        393 \\
 BEL &         5182 &           22 &  GTM &         4687 &          402 &  CIV &       1587 &        406 \\
 MYS &         5174 &           25 &  NPL &         4668 &          386 &  GHA &       1578 &        361 \\
 AUT &         5172 &           31 &  MUS &         4655 &          412 &  BEN &       1457 &        373 \\
 DNK &         5170 &           32 &  IRN &         4645 &          361 &  MOZ &       1366 &        320 \\
 ISR &         5168 &           31 &  PER &         4615 &          383 &  SYC &       1362 &        308 \\
 BGR &         5164 &           33 &  MLT &         4612 &          415 &  KWT &       1178 &        230 \\
 MEX &         5163 &           34 &  FJI &         4601 &          400 &  CMR &       1156 &        222 \\
 EST &         5162 &           35 &  MDA &         4573 &          509 &  CUB &       1102 &        140 \\
 CZE &         5159 &           44 &  KHM &         4528 &          440 &  SDN &       1064 &        134 \\
 THA &         5156 &           45 &  SAU &         4511 &          465 &  MLI &       1007 &        123 \\
 HUN &         5154 &           45 &  SWZ &         4493 &          479 &  BTN &        956 &         36 \\
 CAN &         5148 &           49 &  URY &         4436 &          515 &  AZE &        924 &         41 \\
 SGP &         5145 &           56 &  AND &         4426 &          569 &  NER &        894 &         65 \\
 SWE &         5137 &           66 &  ARM &         4404 &          468 &  BDI &        874 &         83 \\
 RUS &         5119 &           59 &  PRK &         4370 &          589 &  BFA &        830 &         15 \\
 ZAF &         5116 &           73 &  BGD &         4358 &          461 &  TKM &        798 &         35 \\
 LTU &         5110 &           88 &  DOM &         4347 &          636 &  NGA &        762 &         41 \\
 LUX &         5107 &           85 &  MMR &         4323 &          542 &  QAT &        756 &         47 \\
 VNM &         5106 &           84 &  GEO &         4263 &          599 &  CPV &        701 &         74 \\
 HRV &         5102 &           86 &  HND &         4170 &          641 &  VEN &        667 &         99 \\
 PRT &         5099 &          100 &  MNE &         4168 &          624 &  GUY &        635 &         86 \\
 ROU &         5097 &          100 &  KAZ &         4080 &          661 &  YEM &        611 &         91 \\
 IDN &         5083 &          108 &  SYR &         4064 &          869 &  SUR &        576 &         93 \\
 SVK &         5075 &          122 &  UGA &         4051 &          719 &  GRL &        525 &          1 \\
 EGY &         5074 &           90 &  KGZ &         4050 &          787 &  MRT &        493 &          1 \\
 TUR &         5072 &          130 &  ALB &         4034 &          929 &  SOM &        462 &          1 \\
 FIN &         5071 &          128 &  BHR &         3769 &          886 &  DZA &        431 &          1 \\
 SVN &         5067 &          132 &  MDG &         3723 &          877 &  MDV &        400 &         32 \\
 UKR &         5051 &          113 &  BWA &         3656 &          949 &  PNG &        372 &         29 \\
 NZL &         5042 &          113 &  SEN &         3550 &          848 &  GMB &        338 &          8 \\
 IRL &         5040 &          133 &  ISL &         3541 &          931 &  GIN &        305 &          1 \\
 PAN &         5027 &          161 &  UZB &         3536 &          909 &  CAF &        273 &          1 \\
 BLR &         5023 &          166 &  TZA &         3534 &          815 &  COD &        244 &         30 \\
 BRA &         5022 &          166 &  NAM &         3284 &          767 &  COG &        216 &          1 \\
 TUN &         4980 &          193 &  ETH &         3235 &          891 &  AGO &        184 &          1 \\
 GRC &         4964 &          208 &  LAO &         3199 &          818 &  LBY &        156 &          1 \\
 LVA &         4959 &          231 &  ECU &         3173 &          955 &  GAB &        130 &          1 \\
 PAK &         4954 &          207 &  PRY &         2890 &          799 &  LBR &        104 &         27 \\
 CYP &         4951 &          189 &  LSO &         2822 &         1058 &  GNQ &         61 &          1 \\
 LKA &         4930 &          174 &  NIC &         2751 &          757 &  GNB &         49 &          1 \\
 PHL &         4920 &          244 &  JAM &         2704 &          733 &  ERI &         33 &          1 \\
 NOR &         4913 &          232 &  ZWE &         2670 &          850 &  TCD &         15 &          1 \\
 BIH &         4906 &          239 &  TGO &         2554 &          883 &   &           &    \\      
\hline
\end{tabular}
\caption{Table listing the productivity gap and the top product given by the Complexity ranking performed by each country in the database in 2016. The order follows the top product rank and the label indicates the alpha-3 code of the countries.}
\label{tab:list}
\end{table}

\section*{References}

\bibliographystyle{unsrt}
\bibliography{biblio.bib}

\end{document}